\documentstyle[twocolumn,aps,epsf]{revtex}
\begin{document}
\draft
\newcommand{\abst}[1]{\hspace{10mm}\begin{minipage}[t]{155mm}#1\end{minipage}\vspace{5mm}}
%
%
%
\def\etal{{\it et al~}}
%
%

%
\draft
%
%
%
%
\wideabs{
\title{Why relativity allows quantum tunnelling to 'take no time'? }
%
%
\author{D.Sokolovski}
\address{School of Mathematics and Physics,
         Queen's University of Belfast, 
	 Belfast, BT7 1NN, United Kingdom}
\date{\today}
\maketitle
%
%
\begin{abstract}
In quantum tunnelling, what appears an infinitely fast barrier 
traversal can be explained in terms of an Aharonov-like
weak measurement of the tunnelling time,
 in which the role of the pointer is played
by the particle's own coordinate. A relativistic wavepacket
is shown to be reshaped through a series of subluminal shifts which
together produce an anomalous 'superluminal' result.

\end{abstract}
%
%
\pacs{PACS number(s): 03.65.Ta, 73.40.Gk}
}
%
In 1932 MacColl's \cite{McColl} noted that quantum
tunnelling 'takes no time' in the sense that
the tunnelled pulse can arrive in the detector
ahead of its free counterpart, as if it has 
spent no time in the barrier region $b$.  
This effect been observed experimentally
and extensively analysed by several authors
(for a review see \cite{Rev1}).
There is a broad consensus that 
what appears as  
'superluminal' transmission takes place due to a 
reshaping mechanism (see, for example, \cite{Low2}),  
but the role of the tunnelling time in the wavepacket propagation remain a
contentious issue \cite{Rev2}.
One dilemma one encounters when 
 trying to explain the tunnelling speed up is 
whether quantum mechanics allows a particle
to spend anomalously short ($\ll b/c$)
durations in the barrier as was implied in \cite{Rick} or, as 
we argue below, the effect can be achieved
while maintaining, in agreement with special relativity,
 all velocities bounded by the speed of light $c$. 
A suggestion of how this 
may be possible, can be found in the work of  Aharonov 
and co-workers \cite{Ah1,Ah2,Ah3}, who have shown that under 
certain 'weak' conditions a von Neumann meter may produce
readings well away from the region containing all eigenvalues
of the measured variable. In this Letter,
we demonstrate the analogy between wavepacket scattering 
and a weak measurement of the tunnelling time, which, although
effectively all durations involved are subluminal, produces an apparently
'superluminal' result.  
\newline The authors of \cite{Ah1,Ah2,Ah3} considered a von Neumann measurement
of a variable $\mathcal{A}$, $\mathcal{A}$$|\phi_i>=A_i|\phi_i>, 1 \le i \le N, 
\quad N\gg1,\quad 0 \le A_i \le 1$ on a system
which is prepared in a state $|I>=\sum_i a_i |\phi_i>$ and 
after the measurement is found (postselected) in    
another state $|F>=\sum_i b_i |\phi_i>$. After postselection,
the pointer wavefunction in the coordinate representation is given by
\begin{equation} \label{eq:AHAR}
<\tau|M> = \sum_{i=1}^{N} G(\tau-A_i)\eta_i, \quad\quad \eta_i \equiv
b_{i}^{*}a_i
\end{equation}
where $G(\tau)$ is the initial state of the meter. The 'paradox' observed
by Aharonov \emph{et al} is that if the measurement is weak (the width of $G(\tau)$
exceeds the separation between the eigenvalues $A_i$), a careful choice
of $|I>$ and $|F>$ allows one to ensure that $<\tau|M> \approx G(\tau-\alpha)$ 
with an arbitrary 
 shift $\alpha$ which can then be written
as the weak value \cite{Ah1} of the operator $\mathcal{A}$,
\begin{equation} \label{eq:weak}
\alpha=<F|{\mathcal{A}}|I>/<F|I>=\sum_{i=1}^{N}A_i\eta_i/
\sum_{i=1}^{N}\eta_i.
\end{equation}
 Therefore it is possible (although improbable, since $|<F|I>|^2 \ll 1$) that an experiment would produce, for example, a reading 
of $100$ even though all eigenvalues of $\mathcal{A}$ involved 
do not exceed 
$1$ \cite{Ah1}. In \cite{Ah3} this 'peculiar interference effect' 
is summarised 
as destructive interference in the whole "allowed"
($0<\tau<1$) region and the constructive interference of the tails
in the "forbidden" ($\tau \approx 100$) region.
\newline The relevance of the weak measurement
 analysis to wavepacket tunnelling,
where no external meter is involved and no operator is available
for the time spent in the barrier, is not obvious and will 
be established next. 
Classically, the amount of time a  relativistic particle
with a world line $x(s)=(x_0(s),\vec{x}(s))$ spends in a 
 spatial region $\Omega$ is given by 
\begin{equation} \label{eq:func}
t^{cl}_\Omega[x(s)] \equiv \int_{s'}^{s''}ds W_\mu(x)dx^\mu/ds 
= \int \theta_\Omega (\vec{x})dt. 
\end{equation}
where we have put, in the chosen frame of reference,  $W_\mu(x)=(\theta_\Omega (\vec{x}),0,0,0)$ ($\theta_\Omega (\vec{x})\equiv 1$ for $\vec{x}\in \Omega$ and $0$ otherwise).
Quantally, Eq.(\ref{eq:func}) can be used to distinguish between scenarios in which
a particle spends different durations inside $\Omega$.
Propagator for a spin-zero  Klein-Gordon particle in an electromagnetic field $A_\mu(x)$  can be written as a sum over four-dimensional paths parametrised by the fifth parameter $u$,\cite{F1,Sch1} ($\partial_z \equiv \partial/\partial z, \hbar=c=1$)
\begin{equation} \label{eq:KGprop}
g(x,x') = \int_{0}^{\infty} du \exp(-im^2u/2) \int Dx \exp [ -i\int_{0}^{u}
{\mathcal{L}}(x)du]
\end{equation}
where ${\mathcal{L}}(x)=(\partial_u x^\mu)^2/2+e\partial_u x^\mu A_\mu$.
With the help of Eqs.(\ref{eq:func}) and (\ref{eq:KGprop}) the  propagator  
restricted only to those paths 
which spent inside $\Omega$ a duration $\tau$,
$ \int Dx \rightarrow  \int Dx \delta(t^{cl}_\Omega[x(u)]-\tau)$ is seen 
to be given by 
\begin{equation} \label{eq:KGrest}
g(x,x'|\tau) 
= (2\pi)^{-1/2}\int_{-\infty}^{\infty}\exp(iW\tau)
g(x,x'|W)dW
\end{equation}
where $g(x,x'|W)$ is the propagator for the modified electromagnetic field $A_0(x)+W\theta_\Omega (\vec{x}),\vec{A}(x)$.
Analysing the time evolution of $g(x,x'|\tau)$ leads to the clocked 
Klein-Gordon equation ($p_\mu \equiv -i\partial_{x_{\mu}}$)
\begin{eqnarray} \label{eq:KGcl}
\{[i\partial_{t}+i\partial_{\tau}\theta_\Omega(\vec{x})-eA_0]^2-(\vec{p}-e\vec{A})^2-m^2\}\Phi(x|\tau)=0 
\end{eqnarray}
where $\Phi(x|\tau)$ is the (scalar) amplitude that a particle at 
a space-time point $x$ has previously spent a duration $\tau$ in 
the region $\Omega$.
In the case of the Dirac electrons, no path integral is available
 for the spin degrees of freedom. However, one can still define a (4x4 matrix) amplitude $F_P$ for a given space-time path \cite{F1}.
 On application of a field $A(x)$ the phase of $F_P$ is modified, $F^A_P = \exp\{-i\int A_\mu dx^\mu\}F_P$, and restricting the paths as above yields the clocked Dirac equation
\begin{eqnarray} \label{eq:Dcl}
\{[i\partial_{t}+i\partial_{\tau}\theta_\Omega(\vec{x})-eA_0]\gamma_0-(\vec{p}-e\vec{A})\vec{\gamma}-m\}\Phi(x|\tau)=0. 
\end{eqnarray}
Note that like the clocked Schr$\ddot{o}$dinger equation \cite{S3}, both Eqs.(\ref{eq:KGcl}) and
(\ref{eq:Dcl}) are obtained from their original versions by
replacing $i\partial_{t} \rightarrow i\partial_{t}+i\partial_{\tau}\theta_\Omega(\vec{x})$.
\newline To study the durations involved 
in in one-dimensional transmission of a spinless particle across a rectangular
potential barrier $V\theta_{-b/2,b2}(z)$  considered in \cite{Low1}, 
\begin{figure}[ht]
\epsfxsize=8.5cm
\centering\leavevmode\epsfbox{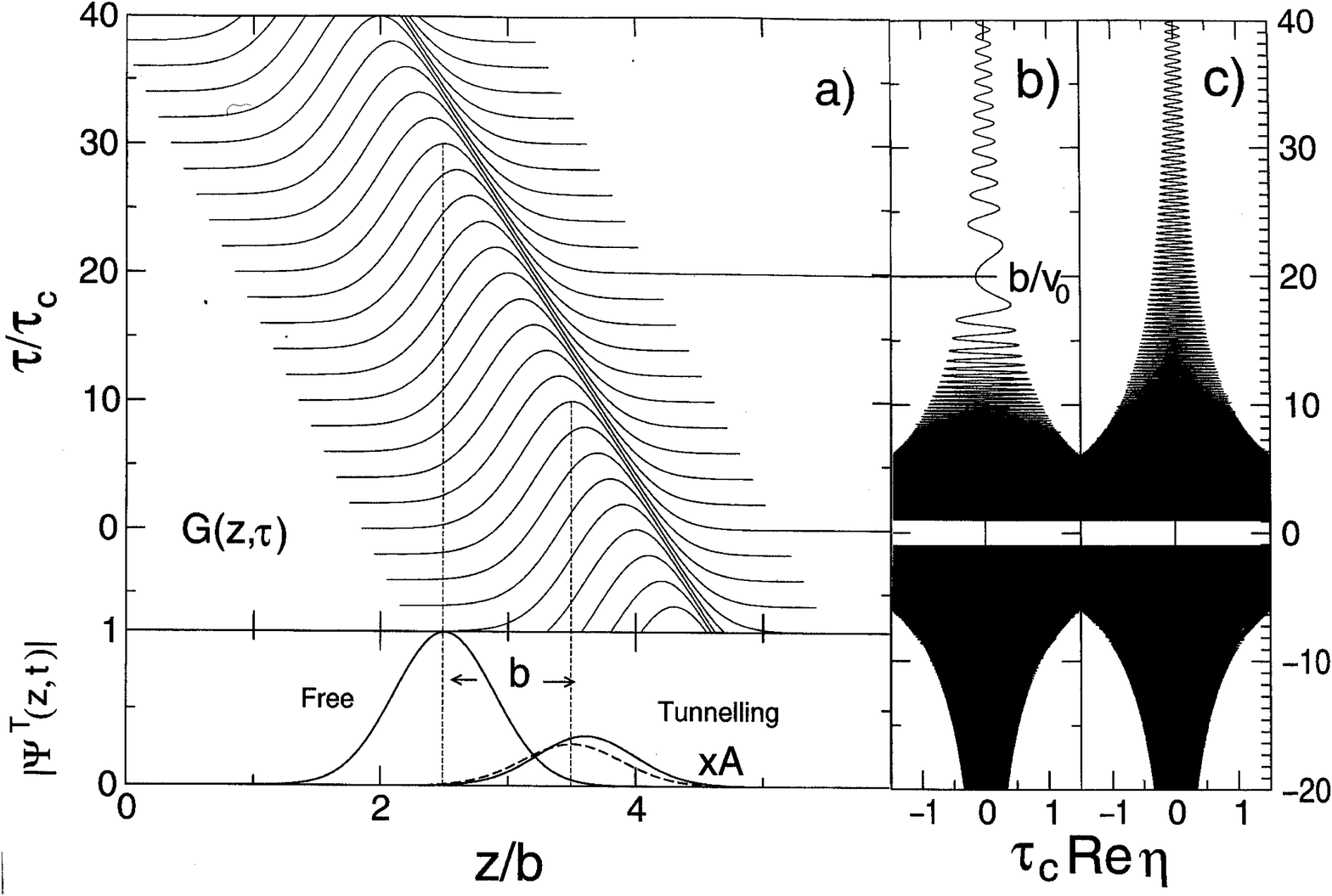}
\vspace{0pt}
\caption{Transmission of a Gaussian wavepacket across the region $[-b/2,b/2]$.
Inset: $|\Psi^{T}(z,t)|$ for free motion and tunnelling ($\times A \equiv
10^{-2} T(p_0,V)$) for
$\epsilon_{0}\tau_{c}/\hbar = 6000$; $\epsilon_{p_{0}}\tau_{c}/\hbar = 6007.5$;
$V\tau_{c}/\hbar=15.$; $\Delta z/b=0.55$; $z_{0}/b=2.5; t/\tau_{c}=100.$;
 $\tau_{c}\equiv
b/c$
Also shown by the dashed line
is the semiclassical result Eq.(\ref{eq:smcl}). (a) the factor $G(z+z_0-b-v_0(t-\tau))$ in Eq.(\ref{eq:WP3});
Real part of  $\eta(p_0,t)$ for (b) free motion, $V=0$; 
 and (c) tunnelling, $V\tau_c/\hbar=15$.}
\label{fig:packs}
\end{figure}

we prepare   
a wavepacket with a mean momentum $p_0$ and a momentum (coordinate) space width $\Delta p$ ($\Delta z$) initially containing only 
 negative frequencies (no antiparticles) 
\begin{eqnarray} \label{eq:inp}
\Psi(z,t=0)=\int A(p)exp(ipz)dp 
=exp(ip_0z)G(z+z_0).
\end{eqnarray}
where $z_0 \gg b/2$ and a suitable  $G(z)$ (e.g., a Gaussian $G(z)=exp(-z^2/\Delta z)$) is chosen to ensure that all momenta in (\ref{eq:inp}) tunnel.
Solving Eq.(\ref{eq:KGcl}) with the initial condition $\Phi(z,t=0|\tau)=\Psi(z,t=0)\delta(\tau)$   
yields, for sufficiently large $z$ and $t$, the traversal time decomposition
of the transmitted part of the wavepacket,
\begin{equation} \label{eq:unfold}
\Phi(z\gg b/2,t|\tau) =  \int dp A(p)exp(ipz-i\epsilon_p t)\eta(p,\tau)
\end{equation}
\begin{equation} \label{eq:eta}
\eta(p,\tau)\equiv(2\pi)^{-1/2}
\exp(-iV\tau) \int_{-\infty}
^{\infty} \exp(iW\tau)T(p,W) dW
\end{equation}
where $\epsilon_p = (p^2+m^2)^{1/2}$ and $T(p,W)$ is the transmission amplitude
for the rectangular potential $W\theta_{-b/2,b2}(z)$.
The amplitude $\Phi(z,t|\tau)$ gives the transmitted field
under the condition that a particle spend in the barrier exactly
a duration $\tau$ 
 and a coherent sum over all such durations produces the transmitted pulse
$\Psi^T(z,t)$
\begin{eqnarray} \label{eq:fold}
\Psi^T(z,t)=\int_{-\infty}^{\infty} \Phi(z,t|\tau)d\tau =\nonumber \\
\int T(p,V)A(p)exp(ipz-i\epsilon_p t)dp
\end{eqnarray}
For tunnelling, ($\epsilon_p-m < V \ll m$), $\Psi^T(x,t|V)$ although greatly
 reduced, lies approximately a distance $b$ ahead of the free ($V=0$) pulse
suggesting that the particle has traversed the barrier
instantaneously (see inset in Fig.1). To see 
 how this speed up is achieved, we consider the 
semiclassical limit $pb, kb \gg 1$ \cite{FT1},  
 replace $T(p,W)$ by the first term in its multiple 
scattering expansion \cite{S2} ($k(W)\equiv[(\epsilon_p - W)^2-m^2]^{1/2}$) 
\begin{eqnarray} \label{eq:trp}
T(p,W) \approx 
 \frac{4pk}{(p+k)^2} 
exp[ikb-ipb],
\end{eqnarray}
 which is permissible both well above, $V\ll\epsilon_p-m$,
and well below, $V\gg\epsilon_p-m$, the barrier,
and evaluate the resulting integrals asymptotically. 
 This yields 
\begin{eqnarray} \label{eq:WP3}
\Psi^T(z,t) \approx \exp(ip_0z-i\epsilon_{p_{0}}t)\nonumber \\
\times \int_{-\infty}^{\infty} G(z+z_0-b-v_0(t-\tau))
\eta(p_0,\tau)d\tau
\end{eqnarray}
together with the saddle point result for $\eta(p_0,\tau)d\tau$ \cite{FT2} 
($\hbar$ and $c$ are reintroduced)
\begin{eqnarray} \label{eq:distr}
\eta(p_{0},\tau) =
\frac{(2\pi\hbar)^{-1/2}4p_{0}b\epsilon_{0}^{3/2}\tau_{c}^{3}}{(\tau^2-\tau_{c}^2)^{1/4}
[p_0b(\tau^2-\tau_{c}^2)^{1/2}+\epsilon_{0}\tau_{c}^2]^2}
\nonumber\\
         \times  exp\{\frac{i}{\hbar}[(\epsilon_p-V)\tau-\epsilon_{0}(\tau^2-\tau_{c}^2)^{1/2}] - i\pi/4\}
\end{eqnarray}
where  $v_0\equiv\partial \epsilon_p/\partial p_0$ and we have introduced $\epsilon_0 \equiv mc^2$ and $\tau_c \equiv b/c$.
Since most of our derivation relies on the analytical properties
of the wavevector $k(W)$, a similar analysis can also be applied
to the Dirac electrons.
\newline  
We proceed with the details of the reshaping mechanism responsible for
the speed up of the tunnelled pulse. 
The first term in the integrand
of Eq.(\ref{eq:WP3}) is the wavepacket translated as a whole
along a classical trajectory which spends inside $[-b/2,b/2]$ a time $\tau$
and has the velocity $v_0$ outside the region (Fig.1a) so that the total
distance covered by the time $t$ is 
\begin{eqnarray} \label{eq:trj}
z-z_0=b+v_0(t-\tau).
\end{eqnarray}
Each wavepacket is weighted 
by the (complex valued) amplitude $\eta(p_{0},\tau)$ for a 
transmitted particle whose momentum is exactly $p_0$ to 
spend the duration $\tau$ in $[-b/2,b/2]$.
It is instructive to consider the case of free motion first.
For $V=0$, the oscillatory amplitude $\eta(p_{0},\tau)$ shown in Fig.1b 
coincides for $\tau \gg \tau_c$ with the non-relativistic 
result of \cite{S2} and has a stationary region around the classical
value $b/v_0$. For $|\tau| < \tau_c$, it rapidly
decays so that, owing to the relativistic nature of Eq.(\ref{eq:KGcl}), the
durations which imply superluminal velocities $v_{bar}=b/\tau$ are explicitly absent from the integral
(\ref{eq:WP3}). For $\tau<-\tau_c$, $\eta(p_{0},\tau)$ becomes oscillatory
again due to the  relativistic paths in  Eq.(\ref{eq:KGprop}) which experience
time reversals associated with creation and annihilation of particle-antiparticle pairs \cite{F1,Sch1}. The frequency of these oscillations is extremely
high and, as one would expect, the processes involving virtual pair production
do not contribute to the propagation. As a result, $\eta(p_{0},\tau)$ 
selects the shape $G(z-z_0-v_0t)$ corresponding
to the classical value $\tau=b/v_0$ 
and, in accordance with classical mechanics, one finds the initial
pulse translated without distortion along the classical path $z(t)=v_0t+z_0$.
\newline In the case of tunnelling, the amplitude $\eta(p_{0},\tau)$
differs from its free counterpart by the factor $\exp(-iV\tau/\hbar)$
in Eq.(\ref{eq:eta}) which destroys
 the stationary region on the real $\tau$-axis (Fig 1c).
There is no preferred duration describing tunnelling and the analogy
with Aharonov's weak measurement is now evident. Just as in Eq.(\ref{eq:AHAR}) a 
sum of Gaussians all shifted by no more than 1 produces a Gaussian 
with an arbitrary large shift $\alpha$, a continuos superposition
of the copies of the initial wavepacket all propagated with purely
subluminal velocities produces a pulse which looks as if it had 
spent a zero time in the barrier. 
The role of the pointer position is played
by the particle's coordinate $z$, and the two distinctive features of a 
weak measurement are in place: the particle is postselected
in a highly unlikely (most particles are reflected) transmitted 
state and the accuracy, which is determined by the coordinate
width of the wavepacket, $\Delta z$ must remain poor 
in order to prevent the particle passing over the barrier, 
 which would destroy the effect. 
Several known properties of the tunnelling transmission can now be clarified.  
Destructive interference between
the maxima and the backward tails of all Gaussians in Fig.1a explains why, as observed in \cite{Low1}, 
the transmitted pulse is formed
from the front part of the initial wavepacket.
Also, like a weak von Neumann meter \cite{Ah2}, quantum tunnelling 
cannot be used to transmit information which cannot be
extracted without a reference to 'superluminal' behaviour.
 Indeed, truncating G(z) after some $z'>> \Delta z$ would guarantee
effective vanishing of $\Psi^T(z,t)$  in Eq.(\ref{eq:WP3}) for $z > ct-z_0$. 
Thus, rather than facilitating faster-than-light information transfer,
a barrrier acts as a filter \cite{Ah2}, rearranging the amplitudes
in the forward tails of all (subluminally shifted) $G$'s in Eq.(\ref{eq:fold})
into a single peak.  
\newline Previous work \cite{Rick,St1,St2,Iann} on relating the tunnelling time
problem to weak measurements focused mostly on interpreting
the 'complex time' \cite{Fert,S3,S4}
\begin{equation}\label{eq:tbar}
\bar{\tau} = \int \tau\eta(p,\tau) d\tau/\int \eta(p,\tau) d\tau
\end{equation}
as the weak value (\ref{eq:weak}) of the projector onto the barrier region,
$\theta_{[-b/2,b/2]}(z)$ \cite{St1,St2} or its time average, $T^{-1}\int_{0}^{T}
\theta_{[-b/2,b/2]}(t)dt$ \cite{Iann}.  
In \cite{Rick} the authors 
applied a low accuracy Salecker-Wigner clock to Dirac electrons
obtaining, effectively, an estimate $v'_{bar} = b/Re\bar{\tau}$
for the mean velocity in the barrier.
For a relativistic particle $v'_{bar}$ may exceed $c$ 
even though $\eta(p_0,\tau)$ has no support
inside $-b/c < \tau < b/c$
and \cite{Rick} questioned the significance
of this result for predicting 'superluminal' transmission of electrons.
The answer 
is that $\bar{\tau}$, which describes
the work of an inaccurate external clock \cite{St1,S3,S4}, cannot, in general,
predict the position of the tunnelled pulse, whose shape
is determined by the oscillatory integral (\ref{eq:fold}) involving
all moments of $\eta(p_0,\tau)$ for a range of particle's
 momenta $p$.
The semiclassical case (\ref{eq:fold})-(\ref{eq:distr}) is special 
because $\eta(p_{0},\tau)$ has an imaginary saddle at 
\begin{eqnarray}
\tau_V = -i\tau_c \frac {(\epsilon_{p_{0}}-V)}{[\epsilon_{0}^2-(\epsilon_{p_{0}}-V)^2]^{1/2}},
\end{eqnarray}
and the superposition of purely
subluminal shifts (\ref{eq:fold}) results for $G$ in an imaginary shift by $\alpha=-i|\tau_V|$.
thus, within our approximation (see inset in Fig.1),
\begin{eqnarray}\label{eq:smcl} 
\Psi^T(z,t) \approx G(z+z_0-b-v_0(t+i|\tau_V|))\nonumber \\
\times T(p_0,V)\exp [i(p_{0}z-\epsilon_{p_{0}}t)/\hbar]
\end{eqnarray}
and $\bar{\tau}\approx-i|\tau_V|+O(\hbar)$ is the single parameter governing the shape of the pulse and, therefore, the speed of transmission.
As the tunnelling speed
up effect itself, anomalously small values of $Re\bar{\tau}$ have their
origin in the oscillatory nature of the traversal
time  amplitude distribution
$\eta(p_{0},\tau)$, but, in the full quantum case,
the condition $\bar{\tau} < b/c$ is not sufficient for the 'superluminal'
transmission to take place and a more extensive analysis based on Eq.(\ref{eq:fold})
must is necessary.
\newline 
In summary, special relativity restricts the range of
 tunnelling times to the subluminal domain,
yet interference
accounts for the 'superluminal' appearance of the transmitted pulse. 
This somewhat counterintuitive  effect is most 
easily explained in terms of Aharonov's weak measurements. 
A barrier reshapes a semiclassical wavepacket
by propagating components
of the incident pulse with purely subluminal velocities
before reassambling them with complex weights determined by
the potential.
Examining the position of the tunnelled particle, one
may estimate the time it has spent in the barrier.
Of necessity, the accuracy of such estimate (determined
by the initial spread in the particle's position)
is so poor that it yields an anomalously small 'forbidden'
value, suggesting that the barrier was traversed almost
instantaneously, just as a weak measurement on a spin
$1/2$ may return the value $100$ \cite{Ah1}. 
This analysis differs from that of 
Refs. \cite{Ah1,Ah2,Ah3}  in that it,
 contrary to the claim
that the traversal time cannot be defined
'intrinsically' \cite{Yam}, 
involves no external pointer.
That the role of such pointer can be played
by the particle's own coordinate,
can be traced back to 
the classical relation  (\ref{eq:trj}) between  the distance and the
time spent in the region,
which below the barrier takes the 
form of weak measurement-like
correlations between the two quantities.  
Finally we note that, given the universal nature of the weak measurement analysis
stressed in \cite{Ah1,Ah2,Ah3,St1}, 
it is unlikely  
that a deeper or more 'classical' explanation could be found
for the apparent superluminality observed in quantum tunnelling.


\begin{thebibliography}{8.}
\bibitem{FT1} We also choose the momentum space width of the wavepacket $\Delta
p$ to be
 sufficiently small, $b \Delta p$, $b\partial k/\partial p_0 \Delta p$, $p_0 \Delta p t \approx 1$ and neglect terms $\approx (\Delta p)^2$.
\bibitem{FT2} The branch of $(\tau^2-\tau_{c}^{2})^{1/2}$ is
chosen to be positve below the right cut in the complex $\tau$-plane with cuts
along
 $[-\infty,\tau_c]$ and $[\tau_c,\infty]$. The contour of intregration
in Eq.(\ref{eq:fold}) runs above the left and below the right cut.
\end{thebibliography}
\end{document}